\documentclass[letterpaper,prl,twocolumn,showpacs]{revtex4}

\usepackage{graphicx}
\usepackage{amsmath}

\newcommand{\eg} {{e.g., }}

\newcommand{\ie} {{i.e., }}
\newcommand{\lmax} {{l_{\rm max}}}

\newcommand{\pd} {\partial}
\newcommand{\po} {p_{\rm o}}

\newcommand{\qc}{q_{\rm c}}
\newcommand{\Qc}{Q_{\rm c}}
\newcommand{\V}{\langle V\rangle}
\newcommand{\vecr} {{\cal R}}
\newcommand{\Vmax}{{V_{\rm max}}}
\newcommand{\Z}{{\cal Z}}
\newcommand{\Zs} {Z_{\rm s}}
\newcommand{\Zv} {Z_{\rm v}}

\begin{document}

%===========================================================
\title{Critical Swelling of Particle-Encapsulating Vesicles}
%===========================================================

\author{Emir Haleva}
\author{Haim Diamant}
\email{hdiamant@tau.ac.il}

\affiliation{School of Chemistry, 
Raymond \& Beverly Sackler Faculty
of Exact Sciences, Tel Aviv University,
Tel Aviv 69978, Israel}

\date{July 7, 2008}

\begin{abstract}
  We consider a ubiquitous scenario where a fluctuating, semipermeable
  vesicle is embedded in solution while enclosing a fixed number of
  solute particles. The swelling with increasing number of particles
  or decreasing concentration of the outer solution exhibits a
  continuous phase transition from a fluctuating state to the
  maximum-volume configuration, whereupon appreciable pressure
  difference and surface tension build up. This criticality is unique
  to particle-encapsulating vesicles, whose volume and inner pressure
  both fluctuate. It implies a universal swelling behavior of such
  vesicles as they approach their limiting volume and osmotic lysis.
\end{abstract}


\maketitle
%------------------------------------------------

Membrane vesicles are fluctuating closed surfaces of 0.1--10 $\mu$m
scale, made of a flexible bilayer of amphiphilic molecules in aqueous
solution.  Serving as a simple model of biological compartments (\eg
red blood cells), they have been one of the most extensively studied
soft-matter systems \cite{Safran1994,Lipowsky1995,Seifert1997}.
Numerous works have addressed the elasticity and statistical mechanics
of the membrane under various constraints, such as area and enclosed
volume (\eg \cite{Helfrich1973,Seifert1997,Lipowsky1991}), or area and
pressure difference across the membrane
\cite{Helfrich1987,Pleiner1990,Gompper1997}, yielding various shapes
and shape transformations.
Actual vesicles are always immersed in solution and thus contain both
solvent (water) and solute. Such biomolecule-encapsulating vesicles
are ubiquitous in cell functions such as signaling and transport
into (endocytosis) and out of (exocytosis) the cell
\cite{MolBio}.  They are also used in various applications as
microreactors or delivery vehicles for cosmetics and drugs (liposomes)
\cite{Lasic1}.

The hydrophobic core of the bilayer membrane hinders permeation of
both water and solute molecules. Over sufficiently short time,
therefore, the vesicle volume is fixed. Yet, while the activation
barrier for water exchange is of the order of 20$T$ ($T$ being the
thermal energy) \cite{Deamer1996}, the barriers faced by the solute
particles are typically much higher due to their size and/or charge,
resulting in membrane permeabilities which are orders of magnitude
lower \cite{Deamer1996}. Moreover, water exchange across the membrane
can be tremendously enhanced (indeed, biologically controlled) via
aquaporin channels, which lower the barrier to below $8T$
\cite{Agre1992}. At sufficiently long times, therefore, most vesicles
are found in a wide semipermeable regime, where water can be
considered as exchanged between the interior and exterior, while the
solute remains trapped inside.  As a result, it has been assumed that
the exterior solution concentration and number of encapsulated
particles determine the vesicle volume in practice
\cite{Helfrich1973,Seifert1997,Lipowsky1991} --- the mean volume
adjusts through water permeation so as to annul the osmotic pressure
difference across the membrane.
%XXX
This scenario has been experimentally verified \cite{Sackmann1991} and
utilized to measure membrane permeabilities of various solutes
\cite{Deamer1996} and create osmotic motors \cite{Sackmann1999}.
Volume fluctuations around the osmotically determined mean value have
been considered as well \cite{Lipowsky2005,Piotto2004}.  However, at
high swelling, as the vesicle approaches its maximum volume, this
volume-adjustment description must break down, and appreciable
pressure difference and surface tension will begin to build up.
Further swelling eventually leads to pore formation and osmotic lysis
\cite{MolBio,Koslov1984,Peterlin2008}.  The change in swelling
behavior,
%XXX
in particular, whether it is a smooth crossover or a sharp transition,
is the focus of the current Letter.

We describe the vesicle as a closed surface composed of $N$ molecules
and having maximum volume $\Vmax\sim a^3N^{3/2}$, $a$ being a
molecular length comparable to the membrane thickness.  It is assumed
that at $\Vmax$ the vesicle has a nonextensive number of
configurations.  The vesicle encloses $Q$ solute particles, which do
not directly interact with the surface other than being trapped inside
it. The vesicle is immersed in a solution of fixed concentration and
temperature, which exerts an outer osmotic pressure $\po$ on the
membrane. Since solvent molecules are exchanged between the interior
and exterior, the vesicle volume is not specified and, hence, neither
are its inner particle concentration and pressure. Thus, the partition
function %XXX
involves integration over all possible volumes,
\begin{equation}
  \Z(T,\po,Q,N)=\int dV \Zv(T,V,N) \Zs(T,V,Q) e^{-\po V/T},
  \label{Z}
\end{equation}
where $\Zv$ and $\Zs$ are the canonical partition functions of the
vesicle and solute particles, respectively, for a given volume $V$.
The thermal energy $T$ is hereafter set to unity. For the solute we
write
\begin{equation}
  \Zs = e^{-Qf(Q/V)},
  \label{Zs}
\end{equation}
where $f$ is the canonical free energy per solute molecule.

A key issue for the highly swollen vesicles studied here is how $\Zv$
behaves as $V$ approaches $\Vmax$. It is shown below that, quite
generally,
\begin{equation}
  \Zv(V\lesssim\Vmax) \sim (\Vmax-V)^{\alpha N},
\label{Zv}
\end{equation}
where $\alpha$ is a coefficient of order unity. This result readily
follows from the two requirements, that (i) the vesicle free energy be
extensive in $N$ for $V<\Vmax$, and (ii) the probability density
function of volumes vanish at $\Vmax$.  In more detail, $\Zv$ is found
by integrating over all surface configurations the factor
$e^{-H[\vecr]}\delta(V-V[\vecr])$, where $V[\vecr]$ is the volume of
configuration $\vecr$, and $H[\vecr]$ its energy (including, \eg
contributions from bending rigidity and surface interactions). For
$V\simeq \Vmax$ one can generally represent the configurations by the
amplitudes $\{u_n\}$ of $N$ normal modes.  (For example, in the case
of nearly spherical vesicles these are spherical harmonics
\cite{Seifert1997,Safran1987,Lebedev1996}.)  One then expands
$V[\vecr]\simeq\Vmax-\sum_n C_n|u_n|^2$.  Assuming that $H$ is
nonsingular at $\Vmax$, and using the integral representation of the
delta function, we get
$\Zv\sim e^{-H(\Vmax)}\int d[u_n]dp\exp[ip(\Vmax-V-\sum C_n|u_n|^2)]$.
Integration over $\{u_n\}$ gives a factor of $p^{-1/2}$ per mode
which, upon integration over $p$, yields Eq.\ (\ref{Zv}) with
$\alpha=1/2$ \cite{ft_alpha}.

\begin{figure}[tbh]
  \vspace{0.75cm} \centerline{\resizebox{0.48\textwidth}{!} 
  {\includegraphics{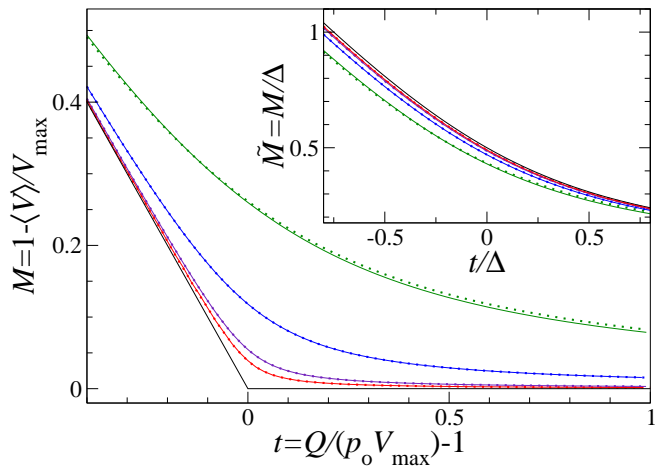}}} \caption[]{(Color online) Order
  parameter as a function of control parameter for an ideal solution
  encapsulated in vesicles of various sizes. Solid curves show the
  mean-field results [Eq.\ (\ref{V})], while dotted curves are
  obtained from numerical integration of the partition function
  [Eq.\ (\ref{Z})]. Datasets from top to bottom (bottom to top in the
  inset) correspond to $\po=1,\ N=30$ (green); $\po=1,\ N=10^3$
  (blue); $\po=5,\ N=10^3$ (indigo); $\po=1,\ N=10^5$ (red); and the
  $N\rightarrow\infty$ limit [Eq.\ (\ref{mlimit}), black]. Only for
  the smallest vesicle size ($N=30$) are the mean-field and numerical
  results distinguishable. Inset shows rescaled data according to Eq.\
  (\ref{mscale}), where the uppermost curve (solid black) is the
  theoretical scaling function. Values of $\po$ are in units of
  $T/a^3$; in all data sets $\alpha=1/2$ and $\Vmax=a^3N^{3/2}$.}
\label{fig1}
\end{figure}

Substituting Eqs.\ (\ref{Zs}) and (\ref{Zv}) in Eq.\ (\ref{Z}) while
specifying the solute free energy $f$, one can perform the integration
over $V$ for given $Q$, $\po$, and $N$. In Fig.\ \ref{fig1} we present
the resulting mean volume, $\V=-\pd\ln\Z/\pd\po$, as a function of $Q$
for an ideal solution, $f(Q/V)=\ln(Q/V)-1$. As $N$ is
increased, $\V$ is seen to approach a
discontinuous first derivative at $Q_{\rm c}=\po\Vmax$.

We now proceed to investigate this criticality analytically for a
general (nonideal) solution. The partition function can be rewritten
as $\Z\sim\int dV e^{-F}$, with
$F = -\alpha N\ln(\Vmax-V) + Qf(Q/V) + \po V$.
Minimizing $F$ with respect to $V$ and applying a first-order
saddle-point approximation (which is equivalent to a mean-field
assumption), we obtain the following equation for $\V$:
\begin{equation}
  Q^2 f'(Q/\V)/\V - \po\V  = \alpha N \V/(\Vmax-\V).
  \label{Vimplicit}
\end{equation}
Expansion in $\Vmax-\V$ yields for our order parameter,
$M=1-\V/\Vmax$,
\begin{eqnarray}
\label{V}
  M &&= \left[\sqrt{[s-t(q)]^2 + 4sg(q)}- s - t(q) \right]
  /(2g(q))\\
\label{mlimit}
  &&\xrightarrow{N\rightarrow\infty} (|t|-t)/(2g),
\end{eqnarray}
where $q=Q/\Vmax$ is the solute concentration at $V=\Vmax$,
$t(\rho)=\rho^2f'(\rho)/\po-1$ is the rescaled difference between the
solute pressure at concentration $\rho$ and the outer pressure
[$t(\rho=q)$ acting as the control parameter of the transition],
$s=\alpha N/(\po\Vmax)\sim N^{-1/2}$, and $g(q)=q^2[2f'(q)+q
f''(q)]/\po>0$ \cite{ft_g}.

Equations (\ref{V}) and (\ref{mlimit}) describe a phase transition at
$q=\qc$ which solves the equation $t(\qc)=0$, \ie for which, if the
volume were equal to $\Vmax$, the inner pressure would just balance
the outer one. The parameter $t(q)$ is related to the actual control
parameter, $Q$ or $q$, via the solute equation of state. 
%XXX
In the ideal-solution example, $f(Q/V)=\ln(Q/V)-1$, the
critical point is at $\qc=\po$ ($\Qc=\po\Vmax$), and we have
$t=Q/\Qc-1$ and $g=1$.  The transition occurs in the region
$|t|\lesssim\Delta=(4sg)^{1/2}\sim a^{-3/2}\po^{-1/2}N^{-1/4}$, along
which $M$ crosses over from finite values to very small ones,
\begin{equation}
  M = 
  \begin{cases}
    -t/g \sim N^0|t|^1 & t \ll -\Delta\\
    \Delta/(2g) \sim N^{-1/4}|t|^0 & |t| \ll \Delta\\
    \Delta^2/(4gt) \sim N^{-1/2}t^{-1} & t \gg \Delta.
  \end{cases}
\label{m}
\end{equation}
In the thermodynamic limit, $N\rightarrow\infty$, this crossover turns
into a sharp corner [Eq.\ (\ref{mlimit})], \ie a discontinuity in
$\partial M/\partial t$.  
%XXX
From Eq.\ (\ref{V}) we find that $M$ follows a scaling law within the
transition region,
\begin{equation}
  M/\Delta = g^{-1}\tilde{M}(t/\Delta), 
  \quad \tilde{M}(x)=(\sqrt{x^2+1}-x)/2,
\label{mscale}
\end{equation}
which is verified in Fig.\ \ref{fig1} (inset).  In addition, we
calculate from Eq.\ (\ref{mscale}) the compressibility,
$\chi = \pd M/\pd\po$,
\begin{equation}
  \chi = (g\po)^{-1}\tilde{\chi}(t/\Delta), 
  \quad \tilde{\chi}(x)=(1-x/\sqrt{x^2+1})/2.
\label{chi}
\end{equation}
Performing the next-order saddle-point calculation (\ie including
fluctuations beyond mean field) yields negligible corrections to $M$,
of order $N^{-3/2}$, $N^{-5/4}$, and $N^{-1}$, respectively, in the
three domains of Eq.\ (\ref{m}). Thus, the mean-field description is
accurate, as is also confirmed by numerical integration of the
partition function (Fig.\ \ref{fig1}).

Equation (\ref{Vimplicit}) (upon division by $\V$) is just the Laplace
law, balancing the pressure difference across the membrane (left-hand
side) with a surface term (right-hand side). We therefore identify the
pressure difference and surface tension as
\begin{eqnarray}
  \Delta p &=& (\alpha N/\Vmax)M^{-1} \sim a^{-3}N^{-1/2}M^{-1},
\nonumber\\
  \sigma &\sim& R\Delta p \sim a^{-2}M^{-1},
\label{Deltap}
\end{eqnarray}
where $R\sim aN^{1/2}$ is the vesicle radius.  Thus, $\Delta p$ and
$\sigma$ change from negligible values below the transition to
appreciable ones above it. Specifically, $a^3\Delta p$ is of order
$N^{-1/2}$, $N^{-1/4}$, and $1$, while $a^2\sigma\sim 1$, $N^{1/4}$, and
$N^{1/2}$, below, at, and above the critical point, respectively.

Note that Eqs.\ (\ref{Z})--(\ref{Zv}), which underly the entire
analysis, contain no microscopic information. The model, therefore, is
purely thermodynamic, in the sense that any specific microscopic model
for the vesicle and encapsulated solution (so long as the vesicle has
a state of maximum volume and negligible entropy) should lead to the
same results. (For example, inclusion of bending rigidity will merely
change the prefactor in Eq.\ (\ref{Zv}).) %XXX 
The invariance to the choice of model implies also that the continuous
transition does not necessarily involve a divergent correlation length
\cite{ft_xi}. We have checked these statements for the specific
example of a nearly spherical envelope of $N$ nodes and fixed total
area $4\pi R_0^2$, enclosing an ideal solution. The vesicle shape is
defined in this case by $R(\theta,\varphi)$, the distance of the
membrane from the center as a function of solid angle, whose deviation
from $R_0$ can be decomposed into spherical harmonics,
$R(\theta,\varphi)=R_0[1+\sum_{l=0}^\lmax
\sum_{m=-l}^l u_{lm} Y_{lm}(\theta,\varphi)]$, where $(\lmax+1)^2=N$.
%Imposing the constraint of fixed total area
%determines one amplitude ($u_{00}$) in terms of the others, and the
%resulting functional for the vesicle volume is given, to quadratic
%order in $u_{lm}$, by $V[u_{lm}]=\Vmax-(R_0^3/4)\sum_{l=1}^\lmax
%\sum_{m=-l}^l (l-1)(l+2) |u_{lm}|^2$. The Hamiltonian is
%$H[u_{lm}]=Q[\ln(Q/V[u_{lm}])-1]+\po V[u_{lm}]$, and 
Integration of the resulting partition function over the amplitudes
$u_{lm}$ within a saddle-point approximation recovers Eqs.\ 
(\ref{V})--(\ref{chi}). 
The correlation function,
%XXX 
$\langle u_{lm}u_{l(-m)}\rangle \sim (M/N)/[l(l+1)-2]$, $l>1$,
exhibits a critical suppression of amplitude but no divergent
correlation length.  Expectedly, this fluctuation spectrum is
identical to that of a spherical membrane with surface tension
$\sigma\sim M^{-1}$, in accord with Eq.\ (\ref{Deltap}).

Equations (\ref{mscale}) and (\ref{chi}) characterize the sharpening
of the transition with increasing system size. If we recast them in
the conventional finite-size scaling form \cite{Barber}, 
$M\sim
R^{-\beta/\nu^*}\tilde{M}(R^{1/\nu^*}t)$ and $\chi\sim
R^{\gamma/\nu^*}\tilde{\chi}(R^{1/\nu^*}t)$, we readily extract
$\beta=1$, $\gamma=0$, and $\nu^*=2$. The values of $\beta$ and
$\gamma$ are consistent with the linear increase of $M$ below the
transition [Eq.\ (\ref{mlimit}) and Fig.\ \ref{fig1}].
Notwithstanding the absence of a divergent correlation length, one can
use $\nu^*$ to define a length scale, $\xi\sim a|t|^{-\nu^*}$, such that
the system lies in the critical domain if $R<\xi$. The divergent $\xi$
does not relate to correlations but to the competition between surface
degrees of freedom ($\sim N$) and three-dimensional ones ($\sim
N^{3/2}$). This competition determines the width of the transition,
$\Delta\sim [N/(\po\Vmax)]^{1/2}$, making it shrink with increasing
$N$. Repeating the analysis for a ring in two dimensions yields a
similar mean-field transition with identical exponents.  We are not
aware of another transition whose mean-field limit has the exponents
found above \cite{ft_Binder}.

The phase transition just characterized is a unique feature of
particle-encapsulating vesicles.  It is a consequence of
the effective inner pressure being dependent on the volume (through
$f(Q/V)$ for fixed $Q$), while the latter fluctuates. This leads to
pressure difference and surface tension which are nonanalytic in $Q$
[Eq.\ (\ref{Deltap})] and a consequent breaking of the equivalence
between the fixed-pressure (or fixed-tension) scenario and that of
fixed $Q$.  Indeed, if the enclosed solution is replaced with a given
inner pressure $p_{\rm i}$ (\ie upon substituting in Eq.\ (\ref{Z})
$Z_{\rm s}=e^{p_{\rm i}V/T}$), it is straightforward to show that
$M=\alpha N [(p_{\rm i}-\po)\Vmax]^{-1}$, in agreement with Eq.\
(\ref{Deltap})
\cite{epje1}. Therefore, in the case of a given pressure difference
(or tension) the vesicle swells gradually with $p_{\rm i}$ (or
$\sigma$) without criticality.  Furthermore, %XXX
replacing the particle-number constraint with a chemical potential 
\cite{ft_mu} is
equivalent (via the solute equation of state) to specifying the inner
pressure. Hence, there is no criticality in the grand-canonical case
either, and the two ensembles are manifestly not equivalent
\cite{epje2}.

Another noteworthy limit is that of a pure solvent outside the
vesicle, $\po\rightarrow 0$, %XXX
where the current analysis yields
$Q_{\rm c}\rightarrow 0$ and $\Delta\rightarrow\infty$, \ie the
transition disappears. %XXX
In this case the swelling of the vesicle toward its maximum volume
occurs already for much smaller particle numbers $Q$, scaling with the
area $N$ rather than the volume \cite{epje2}.

In summary, we have found that membrane vesicles, under rather general
conditions, behave critically as the number of solute particles
inside them is increased or, equivalently, the outer osmotic
pressure is decreased. It should be possible to experimentally observe
this phase transition, \eg by creating vesicles and subsequently
diluting the outer solution in a controlled manner, or by using
isotonic solutions of molecules with different membrane permeabilities
\cite{Peterlin2008}.
We mention three points relating to such experiments.  First, they
should cover such time scales that the vesicle could be considered
permeable to water. This can be sensitively controlled if water
(aquaporin) channels are incorporated in the membrane, yet common
lipid vesicles are also found in this regime over readily accessible
time scales ($\sim 10$ s)
\cite{Sackmann1991,Sackmann1999,Peterlin2008}.
Next we examine the assumption of a sharply defined maximum volume.
One definition of $\Vmax$ would be the volume enclosed by an
unstretchable vesicle of a given area once out-of-plane fluctuations
have completely vanished \cite{ft_nonlinear}. This assumption should
be relaxed when inplane (stretching) fluctuations become comparable to
transverse ones. Since, for a tense membrane, the mean-square
fluctuations of both modes have the same (quadratic) dependence on
wavenumber, this crossover will occur simply when the surface tension,
$\sigma\sim (T/a^2)M^{-1}$ [Eq.\ (\ref{Deltap})], becomes comparable
to the membrane stretching modulus, which is typically of order $10^2$
dyne/cm \cite{Evans1990}.  For $a\sim 1$ nm this happens when
$M\lesssim 10^{-2}$, \ie when the mean volume deviates from $\Vmax$ by
less than 1\%. Thus, we expect the transition from appreciable to
small values of $M$, along with the corresponding scaling behavior, to
be manifest well before stretching becomes important.  Since
lipid membranes can sustain inplane strains of only a few percent
before rupturing
\cite{Koslov1984}, the crossover to stretching-dominated dynamics will
be shortly followed by vesicle lysis \cite{Peterlin2008}.
Third, because of the weak dependence of the transition width on $N$,
$\Delta\sim (a^3\po/T)^{-1/2}N^{-1/4}$, the observed behavior will not
be very sharp. A typical micron-sized vesicle has about $N\sim 10^8$
molecules in its membrane, leading, for a $0.1$M solution, to
$\Delta\sim 10^{-1}$ only. The criticality could be verified,
nonetheless, by checking data collapse according to the scaling law,
Eq.\ (\ref{mscale}).

Thus, our assumptions concerning permeability, maximum volume, and
number of molecules do not rule out an experiment aimed at the
predicted critical swelling. (The suppression of small fluctuations
near the transition, however, may be hard to resolve.) More generally,
this study highlights the qualitative difference between 
semipermeable, particle-encapsulating vesicles and those
having fixed volume or pressure. Since most natural and industrial
vesicles belong to this class, their different behavior should be
taken into account, particularly in cases of high swelling and osmotic
lysis.

%------------------------------------------------
\begin{acknowledgments}
  We thank D.\ Ben-Yaakov and G.\ Haran for helpful conversations. HD
  wishes to thank the Racah Institute of Physics, Hebrew University,
  for its hospitality.  Acknowledgment is made to the Donors of the
  American Chemical Society Petroleum Research Fund for support of
  this research (Grant no.\ 46748-AC6).
\end{acknowledgments}

%------------------------------------------------
% References
%------------------------------------------------

%------------------------------------------------


\begin{thebibliography}{99}
\bibitem{ft_alpha} Thus, provided
  that $N$ is the number of independent modes, we always have
  $\alpha=1/2$. Since accurately relating the number of molecules with
  the number of modes while considering all constraints may not always
  be straightforward, we keep a general value of $\alpha$ in the
  formulation.
\bibitem{ft_g} The function $g(\rho)$ is proportional to the
  derivative of the solute chemical potential with respect to $Q$,
  which is positive if the solution is far from
  condensation.
\bibitem{ft_mu} We do not consider diffusion of particles across the
membrane, but a diffusive contact with a distant buffer, since the
former will entail a vanishing pressure difference.
\bibitem{ft_nonlinear} The derivation of Eq.\ 
  (\ref{Zv}) does not depend on specific normal modes and holds,
  therefore, also for combined transverse and inplane
  fluctuations. One may define $\Vmax$ then as the volume at which
  nonlinear-elastic terms set in.  Yet, since the membrane should
  rupture close to this point, this extension does not seem
  experimentally useful.
\end{thebibliography}
\end{document}